# GDPR Compliant Blockchains – A Systematic Literature Review

AKM Bahalul Haque, AKM Najmul Islam, Sami Hyrynsalmi, Bilal Naqvi, and Kari Smolander

Software Engineering, LENS, LUT University, Lappeenranta, 53850, Finland

Corresponding author: AKM Bahalul Haque (e-mail: najmul.islam@lut.fi )

This work was funded by Business Finland (project 44559/31/2020)

**ABSTRACT** Although blockchain-based digital services promise trust, accountability, and transparency, multiple paradoxes between blockchains and GDPR have been highlighted in the recent literature. Some of the recent literature also proposed possible solutions to these paradoxes. This article aims to conduct a systematic literature review on GDPR compliant blockchains and synthesize the findings. In particular, the goal was to identify 1) the GDPR articles that have been explored in prior literature; 2) the relevant research domains that have been explored, and 3) the research gaps. Our findings synthesized that the blockchains relevant GDPR articles can be categorized into six major groups, namely data deletion and modification (Article 16, 17, and 18), protection by design by default (Article 25), responsibilities of controllers and processors (Article 24, 26, and 28), consent management (Article 7), data processing principles and lawfulness (Article 5,6 and 12), and territorial scope (Article 3). We also found seven research domains where GDPR compliant blockchains have been discussed, which include IoT, financial data, healthcare, personal identity, online data, information governance, and smart city. From our analysis, we have identified a few key research gaps and present a future research direction.

## I. INTRODUCTION

Blockchain is based on a distributed and synchronized digital database for recording information. The database is maintained by a consensus algorithm and stored in multiple nodes [1]. Blockchain and its associated technologies (e.g., smart contracts) promise a new generation of smart services that help establish transparency, accountability, and trust [2].

In 2018, the European Unions' (EU) General Data Protection Regulation (GDPR), which seeks to harmonize data protection laws across the EU states and aims to give back control of data to its owner, came in place [3]. This raised multiple paradoxes between blockchain and GDPR. The challenges arise concerning the implementation of blockchain technology while complying with the GDPR recommendations. For instance, two of the most evident challenges described by the European Parliamentary Research Service (EPRS) are [4, 5]:

i. As per recommendations of the GDPR, there should be at least one data controller. Controller will manage consent of data subject and shall be reachable by the data subjects. However, blockchain is based on decentralization, which would mean replacing a data controller with many different players—this makes the allocation of responsibility and accountability burdensome [6].

ii. GDPR recommends data minimization, purpose limitation, modification, and deletion of data when necessary. However, blockchain is an append-only database that continuously grows as new data is added. Modification and deletion are not recommended in the blockchain to ensure data integrity and trust [7].

These challenges and tensions are multifaceted in nature. There is no single solution that may fit in all business cases. For example, EPRS pointed that "blockchain architects need to be aware of this [challenge] from the outset and make sure that they design their respective use cases in a manner that allows compliance with European data protection law". However, prior research observed that blockchains and data protection regulations are a less discussed topic in the academic literature [8]. Indeed, a few technical solutions have been proposed in prior research for addressing some of the GDPR article requirements [9-11]. While there is a number of literature studies summarizing blockchain research [71-73], its applications [74-76], and even ethical dimensions [77, 78], however, till date, no systematic review exists to summarize the progress on the topic of integrating GDPR and blockchain. As this remains a critical legal and practical question for the adoption of blockchain technology in the European Union, the purpose of this paper is to conduct a systematic literature review on the topic and answer the following research questions.

*RQ1: Which GDPR articles (compliance issues) have been explored for blockchain integration?*

*RQ2: What research areas have been explored about blockchain and GDPR compliance?*

*RQ3: What are the research gaps in the GDPR compliant blockchain domain?*

We conducted our SLR by following the methodology set by Kitchenham and Charters [12]. After running the analysis and applying the inclusion and exclusion criteria, we considered 39 research articles in this SLR. We conducted an in-depth study of these research articles and thematically categorized the GDPR articles that are critical to the blockchain. After that, we summarized possible solutions to tackle the GDPR issues. Moreover, we have also thematically categorized the literature based on the application areas. These led us to identify research gaps that can be filled in future research.

The rest of the paper is structured as follows. After this introductory section, we introduce blockchain technology and GDPR briefly in section 2. Then in section 3, we present the SLR methodology. In section 4, we summarize the findings from the SLR, answer the RQs and provide discussions on the findings. In section 5, we describe the contribution of this paper. Finally, section 6 concludes the paper.

## II. BACKGROUND

### A. BLOCKCHAIN

Blockchain technology has unique characteristics that make it very favorable for many applications [13]. The main features of blockchain are decentralization, consensus mechanism, provenance, immutability, and finality. Decentralization means there is no single most powerful entity that controls the blockchain [14]. The whole system runs on the common agreement of its participants. This common agreement is called consensus [15]. When all the blockchain network participants agree on a transaction, only then the transaction is executed. The consensus mechanism makes the system very trustworthy. Provenance characteristic ensures the traceability of data blocks. In a blockchain network, each block is traceable. If an asset is sold on a blockchain system, it must have detailed construction time of the previous owner's detail saved in its blocks. When a piece of valid information is recorded in a specific block, no one in the network can change or alter it. This immutable nature of blockchain makes it more secure than other methods [16]. Blockchain technology has its application in various sectors, for example, financial, IoT, healthcare, smart city, etc. Blockchain provides transparency and trust among the stakeholders [17] [18].

There are two major types of blockchain: public and private. A public blockchain is spread over a large geographical location. The data is visible to everyone. Although a public blockchain can keep data confidentiality and integrity, it lacks data privacy [19]. Private and consortium blockchain, on the contrary, spread over a specific region fixed by the enterprises. Thus, the data inside this type of blockchain is comparatively restricted [20].

### B. GDPR

GDPR is designed for the EU citizens to control their data and information. GDPR was launched as a replacement of the EU's archaic data protection directive in 1995, later reserved for the UK's information and data protection law in 1998. This was later replaced by the general data protection regulations [21]. The sole aim of GDPR is to make it easier and cheaper for companies to comply with data protection rules [22]. With the rapid growth of digital technologies and an increasing digitalization rate in all sectors like business, healthcare, education, the governments, and citizens are concerned about privacy issues.

Consequently, on May 25, 2018, the new general data protection regulation was launched. It creates a significant impact on all sectors of industries. GDPR focuses on personal data in general. This regulation helps the user to understand how their data is being used. Similarly, the regulations guide the companies on how to manage and process personal data. Any sort of data that can identify the data subject is termed as personal data. It can be the name, phone number, location data, email address, GPS data tied to any identifiable information, or even the IP address [23].

A digital service needs to be compliant with GDPR recommendations, and the service provider should have responsibilities as a controller, collector, and processor [24]. The responsibility of the data collector is to move the data from one service provider to another quickly. The data processor must seek permission from the controller to change, add, and delete individuals' stored personal data and information. The data controller has the authority to share data with third-party organizations or countries. To do that, the controller needs to have a holistic idea about the third party's total data management and processing lifecycle. A GDPR compliant digital service must have data minimization, fairness, accuracy, transparency, and confidentiality [25].

Having discussed the characteristics of blockchains and recommendations of the GDPR, there is a need to harmonize between the two to establish common grounds for developing GDPR compliant blockchains. As a step towards developing GDPR compliant blockchains and understanding state of the art, this SLR was conducted. The focus of the SLR was to investigate the questions identified earlier while considering the fundamentals of the blockchains and principles laid by GDPR.

## III. METHODOLOGY

In this paper, we followed the Systematic Literature Review (SLR) approach to collect primary studies. An SLR is defined as a specific type of review that addresses a focused research theme and detailed representation. In contrast to an unstructured review, an SLR is done by following a predefined procedure, and it is – to some extent – reproducible by other scholars [12]. This study identified the research articles according to the methodology set by Kitchenham and Charters [12]. We selected appropriate keywords for performing the search. After the search results have been found, we applied various parameters and criteria to filter out less relevant articles. After that, we prepared a final list of articles and analyzed them to address the research questions adequately.

### A. LITERATURE SELECTION

One of the most effective methods of selecting literature is setting the required search protocol. Keywords are core components of the search strategy. The keywords are generally produced from the research questions. Moreover, some of the substitutes of the technical words are used. We used Boolean operator "AND". The keywords used include, "Blockchain", "GDPR", "Distributed Ledger Technology", "Data Protection Act". The resultant search strings used for this study are as below.

- ("Blockchain" AND " GDPR")
- ("Distributed Ledger technology" AND "GDPR")
- ("Blockchain" AND "Data Protection Act")

Searches were performed in the title, keywords, and abstracts. The search result was extracted on Nov. 28, 2020. The searched databases include Scopus and IEEE. Scopus is a well-established database that indexes high-quality research articles. This database is widely recognized and researchers across the world use it as a standard. Moreover, we have performed the search on IEEE Xplore Digital library for enlarging the search scope. After the results were extracted, they were filtered through inclusion and exclusion parameters, which are explained in the next section.

### B. EXCLUSION AND INCLUSION CRITERIA

The articles were selected based on an outline set by [12]. The following exclusion and inclusion criteria were applied to extract the final list of articles.

**Exclusion criteria:**
i. Review articles, book chapters, and theses.
ii. Duplicate articles.
iii. Articles that do not discuss compliance issues rather discuss blockchain and GDPR in general.

**Inclusion Criteria:**
i. The study is published in English language.
ii. Full text is available in the digital databases.
iii. The article must be published in a journal or magazine or conference proceeding.
iv. The article reports a case study or proposes a method or a framework on GDPR compliance issues of blockchain.

### C. RESULTS OF THE QUERIES

Initially, 156 articles were found in the Scopus database with the search terms. In addition, 67 articles were found in the IEEE database search. We merged the two databases that resulted in 223 articles. Among them, 46 were duplicates. Therefore, 177 articles remained after removing the duplicates. Among them, 76 articles were found irrelevant and out of scope to our study after reading the titles, and abstracts. After thoroughly reading the full text of the rest of the articles and applying the inclusion and exclusion criteria, 39 articles were considered for this SLR. The article names and sources can be found in in reference section (ref no. [28] – ref no. [66]) Table I shows the initial data extraction based on various keywords.

TABLE I
INITIAL DATA EXTRACTION

| Databases | Keywords | Results |
|---|---|---|
| Scopus | "Blockchain", "GDPR"<br>"Digital Ledger Technology", "GDPR"<br>"Blockchain", "Data protection Act" | 143<br>12<br>1 |
| IEEE | "Blockchain", "GDPR"<br>"Digital Ledger Technology", "GDPR"<br>"Blockchain", "Data Protection Act" | 51<br>9<br>7 |

Figure 1 presents the details of article screening and selection process.

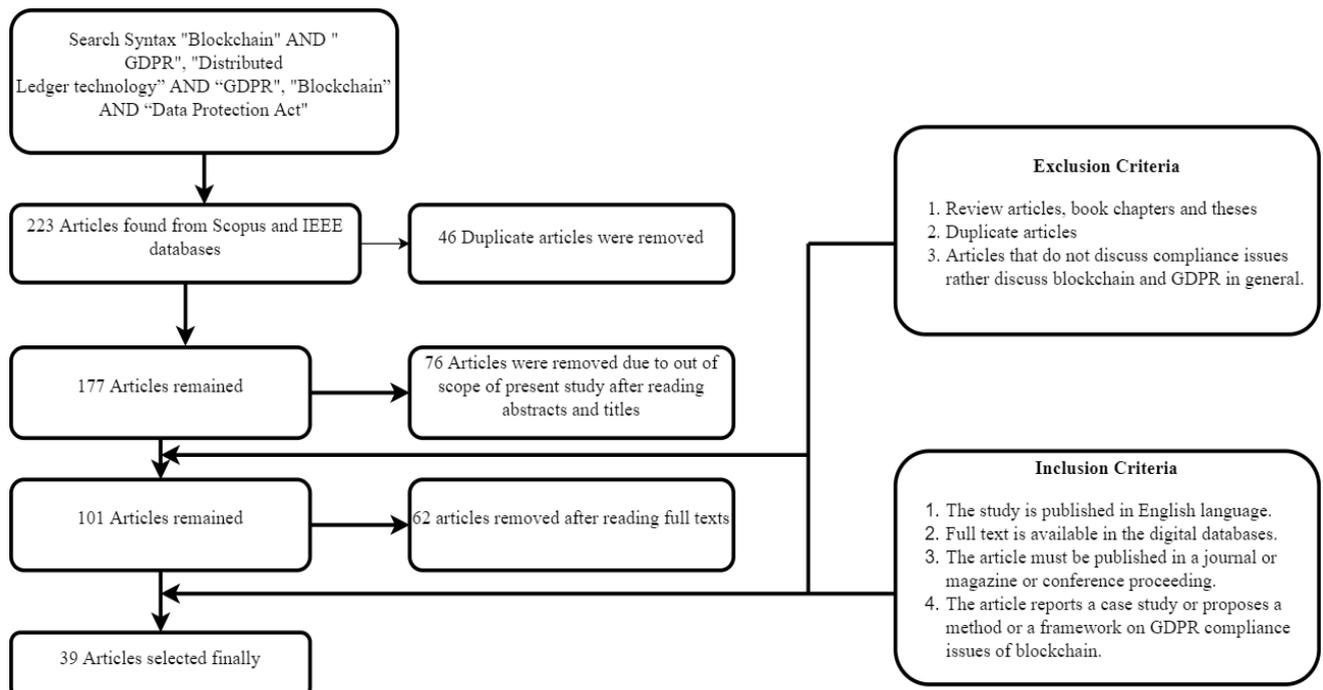

**FIGURE 1.** Article Screening and Selection Process

## IV. RESULTS AND DISCUSSIONS

### A. RESEARCH TREND

Blockchain was first presented in 2008 with the published paper on bitcoin [13]. GDPR, on the other hand, did not start earlier than May 25, 2018, although the discussion about the proposal and legislation started earlier [26, 27].

Research articles published throughout the years are an effective way to observe the research trend and impact (see Figure 2). From the literature search, we observe that articles regarding blockchain and GDPR first began in 2017. Therefore, it is a relatively new research area, and the number of papers keeps increasing with time. Various types of articles have been published over the years. We observed that in total there are 24 conference articles, 13 journal articles, and 2 magazine articles. Table II represents the number of articles and publication years. We observed that there are 24 conference articles, 13 journal articles, and 2 magazine articles considered in this study.

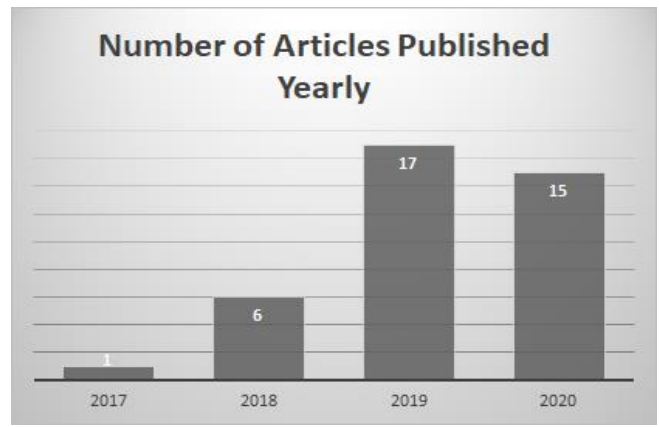

**FIGURE 2.** Publication trend over the years

The number of articles published by various publishers is shown in Table III. We observed that IEEE, ACM, Elsevier, and Springer are the main publishers in GDPR compliant blockchain research.

TABLE II
NUMBER OF ARTICLES PUBLISHED BY VARIOUS PUBLISHERS

| Article Type | 2017 | 2018 | 2019 | 2020 | Total |
|---|---|---|---|---|---|
| Conference | 1 | 5 | 12 | 6 | 24 |
| Magazines |  | 1 |  | 1 | 2 |
| Journal |  |  | 5 | 8 | 13 |
| Total | 1 | 6 | 17 | 15 | 39 |

TABLE III
NUMBER OF ARTICLES PUBLISHED BY VARIOUS PUBLISHERS

| Publishers | Count |
|---|---|
| ACM | 5 |
| Elsevier Ltd | 3 |
| Emerald Group Publishing Ltd. | 1 |
| IEEE | 20 |
| MDPI AG | 1 |
| Springer | 8 |

| Taylor and Francis Ltd. | 1 |
|---|---|
| Total | 39 |

The authors who published the research papers included in this SLR are affiliated with various countries (see Table IV). We observed that most of the affiliations were from the United Kingdom (UK) and EU regions. This is no surprise since GDPR is designed and implemented in the EU regions. However, we also observed some studies from the Asian countries such as China, South Korea, Malaysia, and Taiwan. Though smaller in number, studies have been conducted in North and South American regions as well.

TABLE IV
NUMBER OF COUNTRY-SPECIFIC AFFILIATIONS

| Country | No. of Affiliations |
|---|---|
| Italy | 4 |
| Hungary | 2 |
| Denmark | 4 |
| UK | 24 |
| S. Korea | 10 |
| Germany | 1 |
| Romania | 7 |
| France | 8 |
| Greece | 12 |
| Netherlands | 4 |
| Switzerland | 2 |
| Taiwan | 3 |
| Canada | 4 |
| Malaysia | 4 |
| Portugal | 5 |
| China | 8 |
| Norway | 7 |
| Austria | 10 |
| Australia | 3 |
| USA | 2 |
| Turkey | 2 |
| Poland | 2 |
| Brazil | 3 |
| India | 2 |
| Luxembourg | 1 |

### B. ADDRESSING RQ1

The first research question considered for this study was '*Which GDPR articles (compliance issues) have been explored for Blockchain integration?*' The GDPR articles that have been focused on in prior studies and their proposed compliance approaches are outlined as follows, however, these have been identified in Figure 3 and briefly presented in Table V.

#### 1) DATA DELETION AND MODIFICATION (ARTICLE 16, 17 & 18)

One of the most important and common contradictions found in prior literature is about Article 17 of GDPR that states about "right to be forgotten". That means the concerned organizations should delete the user data if the user requests for it. Since information inside the blockchain cannot be removed, it directly contradicts with Article 17 [28–33]. Technically, the same goes for Article 16 (right to rectification) also since data cannot be edited either [29], [34].

There are three methods proposed to approach this issue. The first and perhaps the most used approach is storing the data in off-chain. It means the original data is stored outside the blockchain and its instance is stored inside the blockchain. The two data are linked with hash and other identifiable information. A specific key management protocol is used, which can be named logical deletion [34, 35]. Logical deletion is performed by destroying the link between the off-chain and on-chain data or removing the private key. The second approach to tackle the problem is to define a consensus mechanism that would delete a block from the chain [36]. A blockchain consensus algorithm is used to add new blocks to the chain after verification. Similar to this, the authors suggested whether a consensus can be designed to delete a block from a blockchain so that each node or the controlling authority can be aware of it and provide consent to it. Finally, smart contract-based solutions are also approached as possible solutions [37], [38]. Smart contracts can execute tasks under certain conditions. In this case, upon user consent, smart contracts can do the task of deleting or modifying the data [35], [39], [40]. Since smart contracts work under specific conditions, user confirmation can be used as that condition.

#### 2) RESPONSIBILITIES OF CONTROLLERS AND PROCESSORS (ARTICLE 24, 26, AND 28)

Article 4 states the definition of personal data, controller, processor, pseudonymization, etc. In turn, articles 24 and 28 describe data controllers' and processors' roles and responsibilities. Blockchain has a specific problem with identifying these roles since there is no centralized authority to control all the nodes [30], [31], [41]. Several approaches have been proposed in prior literature, such as defining the participating nodes as controllers [42], [43], data providers as controllers and miners as processors [44], actors who are able to execute a transaction and add a block as the controllers, joint controllers for federated blockchain [45], and developers as processors for smart contracts [46], [47], [48], [49].

#### 3) PROTECTION AND PRIVACY BY DESIGN (ARTICLE 25)

Prior works highlighted that some of the compliance issues related to privacy and protection by design exist by default in blockchain [36]–[39]. Therefore, these studies viewed that blockchain services are compliant with article 25 of GDPR [30], [47]. Blockchain provides data immutability, confidentiality, and integrity by default. Therefore, subjects' data remains protected against any unauthorized access, data alteration or change, and integrity compromise. Moreover, blockchain uses cryptographic hash functions that make the block data immutable.

#### 4) CONSENT MANAGEMENT (ARTICLE 7)

Article 7 of GDPR emphasizes consent management conditions. The controller's responsibility regarding the data processing upon the user's permission is discussed [31]. The user has the right to give permissions, withdraw consent for processing and storage [31], [51]. Users' data cannot be stored and processed without their consent [52]. As mentioned in the previous section, there are no fixed controllers or processors in the blockchain. For this reason, it is difficult to manage the user's consent to process their data. Saglam et al. [53] discussed that smart contracts can be an effective tool for consent management. The authors highlighted that the compliance codes for consents can be encoded into smart contracts and stored in the blockchain. The contracts can later be used as proof for personal data processing by controllers.

### 5) DATA PROCESSING PRINCIPLES AND LAWFULNESS (ARTICLE 5, 6 AND 12)

Articles 5 and 6 focus on data processing nature, especially mentioning confidentiality, transparency, integrity, and truthfulness of the minimized personal data. Blockchain data is distributed all over the network nodes. Each block uses the hash value of the previous block. This process goes on as long as the chain is not destroyed. So, technically the block data is automatically processed as long as the chain is active [45,53]. On the other hand, GDPR recommends the reduction of automatic data processing [55], [56]. To tackle this, Zemler et al. [29] and Freund et al. [54] highlighted that personal data collected should be as minimal as possible, and processing should be within a limited scope.

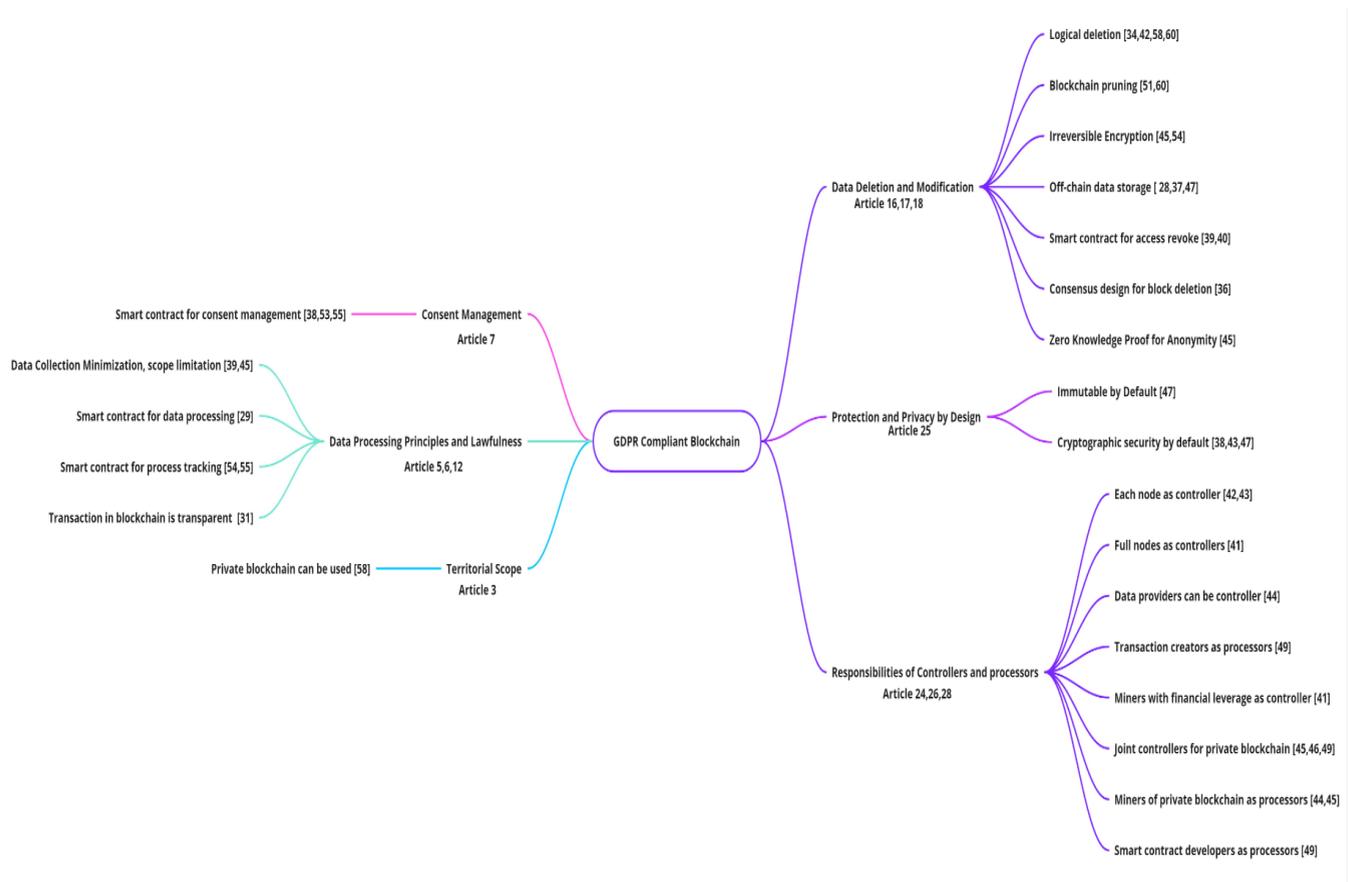

FIGURE 3. Proposed solutions for GDPR compliance

### 6) TERRITORIAL SCOPE (ARTICLE 3)

Article 3 is about controlling the user data from being processed and stored outside the geographical area of the EU. In the case of the public blockchain, it is difficult since the nodes exist worldwide. The situation is different in the case of private and federated blockchain since the nodes are spread over a fixed region [45], [58], [48]. Therefore, private and federated blockchains are preferred to comply with the territorial scope. Besides, the authors discussed the impact of appropriate security measures while data processing and data transfer from one region to another [55].

TABLE V
ARTICLES DISCUSSED AND PROPOSED SOLUTIONS IN PRIOR LITERATURE

| Categories | GDPR Articles and References | Discussed Compliance Issues and Proposed Compliance Approaches |
|---|---|---|
| Data deletion and modification | Article 17 [28-34,59,54,46,47,51,37,42,58,48,38,60,35,36,39,40,44,54,61,53] | • Consensus design for deleting requested block [36].<br>• Integrating compliance code in blockchain design.<br>• Introducing protection by design [38,35,36,44,53].<br>• Smart contract-based solution using non-PBFT (Practical Byzantine Fault Tolerance) consensus in consortium blockchain platform [39].<br>• Controller divides the data into personal and non-personal data [38].<br>• Personal data stored in a local database (off-chain) [38].<br>• Original data can be stored in the off chain and the proof of existence can be stored on-chain using a hash [28,47,37,42,35].<br>• Key management protocols can be used that will be linked with the main chain with hashes. [59,51,37,44].<br>• Link destroyed in the case of right to be forgotten is applied [42].<br>• Destroying the encryption key would work as a logical deletion [34,58,60].<br>• Blockchain Pruning [51,60].<br>• Smart contract for monitoring access revocation [39,40].<br>• Access revocation by using irreversible encryption [54,45]<br>• Recommended use of zero-knowledge proof to make data anonymous [45] |
| | Article 16 [29,34,51] | • Off-chain data storage can be used [28,47].<br>• Data outside blockchain can be rectified/edited normally [48,38].<br>• Blockchain Pruning [51]. |
| | Article 18 [34] | • Smart contracts for consent management [38,62]<br>• Focuses on rights restriction [59]. |
| Responsibilities of controllers and processors | Article 24 [30,45,47,42,48,38,40,41,43,44,49,57] | • Each node as a controller [42,43];<br>• The responsible person who provides data can act as a data controller [44];<br>• Controller and processor roles are defined separately [38].<br>• Controllers share data with processor [38];<br>• Persons with permission to create transactions and update ledger can act as processor [49].<br>• Data controller manages separate data storage as off chain [42].<br>• For bitcoin collective users of full nodes act as controller [41]<br>• Miners having financial capability of enforcing policies can act as controller [41]. |
| | Article 26 [46,49] | • Use of private/permission blockchain for joint controller [46].<br>• Federated blockchain can act as joint controllers [45,49]. |
| | Article 28 [30,45,42,48,38,44,49] | • Controller and processor roles are defined separately [38].<br>• Smart contract developers can act as data processors [49].<br>• Data Separated to personal and non-personal data by controller [38].<br>• Permissioned blockchain-miners are data processors [45,44] |
| Protection and Privacy by Design | Article 25 [30,31,59,47,35,43,44,49,50,63,49,61] | • Immutability nature of blockchain helps with privacy by design [47].<br>• Default attributes make blockchain compliance [44].<br>• Complies with privacy by design since it uses a cryptographic approach by default [47,43,38].<br>• Data security by default using blockchain [49,50].<br>• Proposed privacy by design focused GDPR compliant model for IoT based systems used for navigation [63].<br>• Private blockchain helps implement security measures for cross border data transmission.<br>• Privacy awareness might not comply with blockchain [64]. |
| Consent Management | Article 7 [31,51,38,52,53] | • Need Confirmation of Data Subjects Consent [31,51]<br>• Consent Collection technique must be Distinguishable.<br>• The smart contracts are proposed for consent management [38,53,55].<br>• Conditions of consent and other compliances encoded into smart contract [53,55]. |
| Data Processing Principles and lawfulness | Article 5 [29,45,55,56] | • Smart contracts are used for ensuring legitimate data processes [29].<br>• reduce automatic data processing without user consent [55,56]<br>• Data minimization and scope limitation [29,45] |
| | Article 6 [54,55,56] | • Smart contract for users' data usage audit and tracking [54,55] |
| | Article 12 [31] | • Blockchain based system for data subject's verifiability and right management [31]. |

| Territorial Scope | Article 3 [45,58,53,57] | • Private/Permissioned blockchain can be used [58]. <br> • Permissioned blockchain spreads over a limited geographical region [55]. |
|---|---|---|

### C. ADDRESSING RQ2

The second research question considered in the study was *'What research areas have been explored about blockchain and GDPR compliance?'* The research areas explored about blockchain and GDPR are outlined in the next subsections. We present a summary of the findings in Table VI, and different solutions are presented graphically in Figure 4.

1) INTERNET OF THINGS (IoT)

IoT is the most widely explored research area. We observed from our literature review that blockchain is often integrated with IoT for ensuring security and data integrity. IoT is used in various domains such as healthcare, industrial applications, and smart grid. IoT services need to collect and process versatile personal data to function [64]. Personal data collected by IoT falls under the scope of GDPR. One of the approaches is integrating data protection mechanisms in IoT ecosystem design (Article 25) using blockchain [63]. This approach ensures the integrity, confidentiality, and transparency of data. Another approach proposed is the use of smart contracts [62]. The smart contracts can be used for a compliance audit, verification purposes, data subjects consent management, and compliance history management.

2) HEALTHCARE

Healthcare data comprises a patient's medical history [58]. Therefore, healthcare data is considered as a special category of data according to article 9 of GDPR. Prior literature discussed secure data exchange enabled, trustworthy, traceable, and decentralized data management models [29,57] for healthcare data storage and transaction. Bayle et al. [42], Stan et al. [58] and Zheng et al. [65] discussed the off-chain storage architecture of healthcare data management. Instead of storing the original data in the blockchain, only the proof of existence is stored in on-chain. Bayle et al. [42] and Stan et al. [58] also proposed the mechanism of consent revoking. The link between, on-chain and off-chain can be destroyed for the data deletion upon consent deletion.

In addition, Stan et al. [58] discussed the geographical location of data storage issues. Zemler et al. [29] analyzed the compliance issues for healthcare sectors and discussed the confidentiality, integrity, and trustworthiness of the data that the blockchain system provides. Mohammadi et al. [66] proposed to include a separate privacy-enhancing layer with their proposed GDPR compliant model. Wyciślik et al. [33] highlighted medical documents tracing and auditing systems using blockchain.

3) PERSONAL IDENTITY

An identity management system is needed in most digital services. Identity management systems deal with mostly personal data. Since blockchain provides a tamper-proof mechanism to store personal data, it is an effective and suitable choice. Due to the immutable nature, the data is supposed to be confidential. It helps to build trust and transparency in business and transactions. Giannopoulou [43] and Sim et al. [31] discussed the existing blockchain-based identity management systems and their issues for GDPR compliance. The authors have highlighted compliance issues regarding data modification and deletion. Schmelz et al. [50] presented GDPR compliance issues regarding protecting data subject's rights and data security. Similarly, Pedrosa et al. [56] discussed the data subject's consent management and data minimization. Kondova et al. [46] and Damian et al. [37] proposed GDPR compliant identity management approaches. Damian et al. [37] proposed a GDPR compliant approach that uses ArtChain and PrivateSky protocols and focuses on using on-chain and off-chain architectures. Similarly, Al-Zaben et al. [38] and Truong et al. [34] proposed on-chain and off-chain data storage for personal data management. Truong et al. [34] highlighted using smart contracts for compliance design along with consent management. Jambert et al. [44] emphasized identifying GDPR roles, responsibilities and ensuring privacy by design for personal identity management.

4) ONLINE DATA

Under this category, we have included social network data, e-commerce data, cloud-based data, and other online data. In most cases, consent management is weak in social network websites, but these sites collect a lot of personal data. For this reason, consent collection and management is prioritized. The use of private blockchain is also recommended to have better control and management [52]. Varghese et al. [28] discussed edge marketplace. In case of compliance issues, off-chain storage is recommended. Aujla et al. [55] recommended using smart contracts and blockchain for consent management and data movement tracing in cloud-based data compliance framework. Similar to what we previously mentioned, the use of blockchain technology causes compliance issues regarding the difficulty in role identification, data modification, and deletion. Saglam et al. [53] proposed the use of smart contract and off-chain data storage techniques. Smart contracts can be used for consent management as mentioned in previous sections as well [39][67].

5) FINANCIAL DATA

This category of data includes bank data, insurance data, and cryptocurrency-related compliance issues. In the case of bank-related data, Ma et al. [35] focused on privacy-oriented banking data management. The authors also discussed issues and opportunities of adopting a decentralized environment. Billard et al. [63] highlighted the consent collection and management system for customer's personal data. Cryptocurrencies like Bitcoin are public blockchains. Personal identity in most cryptocurrencies is pseudonymized. Therefore, it falls under GDPR compliance issues. Identifying data collectors and processors are the compliance issues related to bitcoin [41].

6) INFORMATION GOVERNANCE

This is crucial for modern urbanization and organization management. Dutta et al. [45] and Hofman et al. [48] discussed information governance and its compliance problems with GDPR. The authors discussed the roles and applicability of GDR compliant blockchain in information governance. Dutta et al. [45] proposed network administrators and miners as controllers and processors, respectively. The authors have also highlighted smart contract developers as the data controller and the contract itself as the data processor. In addition, for making the data anonymous, the authors discussed zero-knowledge proof.

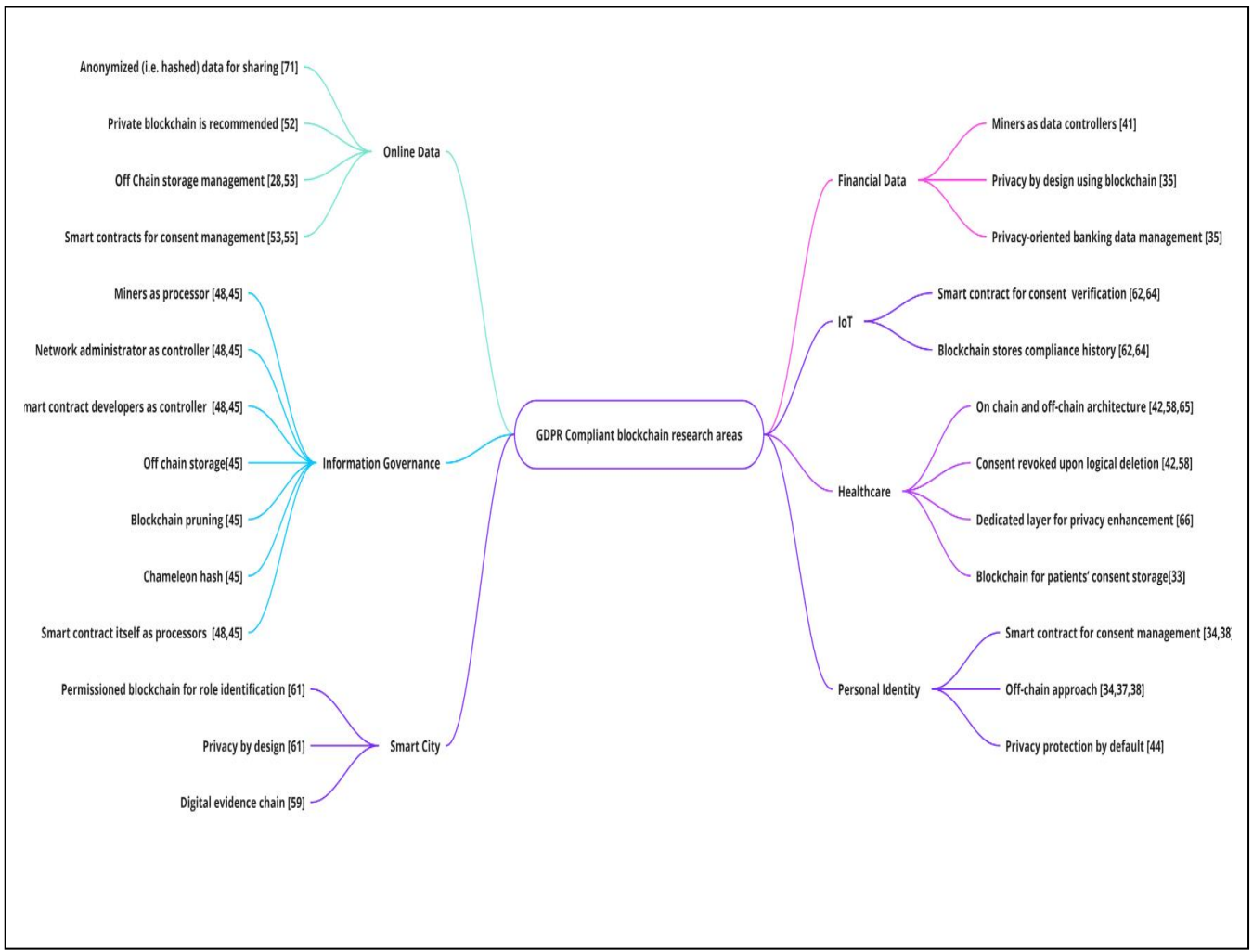

FIGURE 4. Research domains and proposed solution approaches

## 7) SMART CITY

The smart city is the epitome of rapid modern urbanization. Citizens' personal data collection, storage, and transfer happen throughout its network. Ramos et al. [64] and Makhdoom et al. [61] discussed GDPR compliance issues with blockchain in terms of difficulty in determining roles of controller and processor, data deletion, and modification difficulty, etc. The authors also opined that permissioned blockchain makes it relatively easy for determining roles whereas permission-less blockchain makes it challenging.

Makhdoom et al. [61] emphasized privacy by design and consent management. Surveillance systems collect various types of personal data for identifying persons and their history. For this reason, this category of data is crucial to be considered for compliance issues. Asghar et al. [59] discussed surveillance data collection, types of data, compliance issues, and proposed solutions. The authors have proposed the use of off-chain to store personal data with a view to solve the data deletion, modification-related compliance issues. Only the proof of existence shall be stored inside the blockchain [68-70].

TABLE VI
THE RESEARCH DOMAINS DISCUSSED IN PRIOR LITERATURE

| Categories | Research Areas and References | Discussed Compliance Issues and Proposed Compliance Approaches |
|---|---|---|
| IoT | IoT [63,64,62] | <ul><li>Integrating privacy by design in the IoT ecosystem using blockchain [64].</li><li>Focused on privacy by design [63].</li><li>Smart contract for consent verification [64,62].</li><li>Blockchain stores compliance history [64,62].</li></ul> |
| Healthcare | Healthcare [29,33, 42,58,57,71,65,66] | <ul><li>On chain and off-chain architecture proposed [42,58,65].</li><li>Blockchain based identity management [29];</li><li>Data processors and controller roles not identified; data confidentiality, integrity, transparency is discussed [29].</li><li>Consent revoked upon destroying the communication link between on chain and off-chain storage [42,58];</li><li>Focused on the geographic region of data storage [58].</li><li>Focuses on data subjects' right [44]</li><li>Proposed GDPR compliant tool that includes dedicated layer for privacy enhancement [66].</li><li>GDPR compliant solution focuses on secure data exchange [44]</li><li>Solution focuses on trustworthy, traceable and decentralized data management model [44].</li><li>Patient's clinical documents tracing and auditing system [33]</li><li>Blockchain for storing documents and managing patients' consent [33].</li></ul> |
| Personal Identity | Identity Management [31,54,46,37,43,50] | <ul><li>Uses ArtChain and PrivateSky privacy preserving protocols [37,43],</li><li>Focuses on data subjects' right, consent management system including data minimization, on chain and off chain [37,43],</li><li>Discussed noncompliance issues i.e., data deletion, role defining, modification of data etc. [31,43],</li><li>Discussed current methodologies of identity management [43].</li><li>GDPR compatibility with blockchain is discussed for identity management [43].</li><li>Proposed GDPR compliance blockchain for protecting subjects' right, data security with encryption using blockchain [50];</li><li>GDPR compliance architecture design and defining various roles [46].</li></ul> |
| | Personal Data Management in General [34,38,44] | <ul><li>Blockchain based personal data management,</li><li>Smart contract is used to code the GDPR guidelines such as consent management [34,38]</li><li>Personal data is stored off chain in traditional databases [34,38];</li><li>On-chain and off-chain data used for data storage [38].</li><li>Focused on identifying data controller and processor roles in blockchain, privacy by design ensured by several approaches [44];</li><li>Blockchain's storage limitations, weak privacy [36];</li></ul> |
| Information Governance | Information Governance [48,45] | <ul><li>Discussed Roles and applicability of blockchain in Information Governance [48,45]</li><li>Focused on GDPR compliant blockchain design [45].</li><li>Compliance design with Article 17 (off chain, pruning, smart contract, chameleon</li></ul> |
| | | hash) [45]. <br> • Network administrator as controller and miners as processor for permissioned blockchain [48,45] <br> • Smart contract developers as controller and contract itself as processor [48,45]. <br> • Zero knowledge proof makes data anonymous [45]. |
| Online Data | Social Network [52] | • Emphasized on data subjects' consent collection management [52]. <br> • Recommended approach of using private blockchain [52]. |
| | E-Commerce [28] | • On-chain and off-chain data storage [28]. |
| | Cloud based Data. [55,71] | • Cloud based data compliance with GDPR is focused [55]. <br> • Proposed a compliance model framework [55]. <br> • Smart contract and blockchain used for consent management, and data movement trace [55]. <br> • Personal data can be hashed for anonymizing [71]. |
| Financial Data | Bank [35] | • Focused on privacy-oriented banking data management [35]. <br> • Discussed issues and opportunities of adopting decentralized environment [35]; <br> • Discussed privacy by design, right to be forgotten [35]. |
| | Insurance [63] | • Insurance companies collect customer's personal data [63]. <br> • Focuses on user consent collection [63]; |
| | Bitcoin [41] | • Some of the Bitcoin characteristics are non-compliant with GDPR [41]. <br> • Personal identity in bitcoin is pseudonymized [41]. <br> • Collective users of full nodes and specific miners having capability of enforcing policies can act as controller [41]. |
| Smart City | Smart city [49,61] | • Difficulty in determining roles of controller and processor is discussed [49,61]. <br> • Permissioned blockchain makes it easy for determining roles; and permission less blockchain makes it tough [61]. <br> • Data retention period makes it a concern for GDPR Article 17 [49,61]. <br> • Privacy by design, consent management is discussed [61]. |
| | Surveillance System [59] | • Surveillance creates visual personal data [59]. <br> • Data protection by design for surveillance data using blockchain [59]. <br> • On-chain and off-chain storage is proposed [59]. <br> • Off-chain data facilitates more data storage, data deletion, and modification. <br> • On-chain storage stored hash value [59]. |

## D. ADDRESSING RQ3

The third and final research question considered was '*What are the research gaps in the GDPR compliant blockchain domain?*' In the previous sections, we presented the GDPR articles that have been considered, their probable solution approach, and the focused research areas. We also observed that GDPR compliant blockchain is a relatively new research topic. After a careful analysis of the literature considered for the study, the following are some major research gaps that can be explored in the future.

    i. First, our analysis shows various domains that have been explored for blockchain and GDPR related compliance issues. Based on our observations we have not found any article that discusses the issues in the industrial domain. Blockchain is a very widely used technology in the industrial sector. Therefore, compliance issues in the industrial domain can be useful research directions for GDPR and industrial data compliance.

    ii. Second, we observed limited research on smart cities. Smart homes and smart buildings are essential components of future smart cities. Both of these infrastructures deal with personal data and the use of blockchain [67], [68]. For this reason, data privacy and protection regulation compliance are very important. Extensive research and compliance mechanism framework should be very impactful and highly recommended.

    iii. Third, data minimization in blockchain, along with automatic data processing is a major contradiction with the GDPR recommendations [29, 54-56]. Blockchain uses the previous block's hash and continues towards the next blocks. For this reason, the user data entered once, is being processed every time a new block is added. We found only a limited number of research studies that discuss and propose solutions to this problem. Therefore, future research in this direction would be useful.

    iv. Fourth, expanding the research scope for private blockchain. GDPR is perhaps suitable for centralized architecture since this type of architecture is ideally suitable for defining data controllers and processors (Article 24 and 28 respectively). Private blockchain shows a more

- v. centralized framework compared to the public blockchain. It is already proposed that private blockchain miners can act as processors [45]. Moreover, private blockchain can limit the territorial scope. For this reason, further research in private blockchain can be beneficial for mitigating the compliance issues between blockchain and GDPR.
- v. Fifth, blockchain is used for various sectors related to education [69], [70]. Educational institutions deal with the personal data of students, teachers, and other staff. From our SLR, we did not find a study that addresses the data privacy regulations related to the use of blockchain in educational institutions. Therefore, GDPR compliant blockchain in the domain of education is one potential area of future research and use case design.
- vi. Finally, the privacy by design approach is discussed in some of the literature included in this SLR. Privacy by design is one of the most important additions of GDPR. Therefore, it needs to be explored in more detail. In addition to that, the inherent attributes of blockchain that comply with these regulations can be explored elaborately so that GDPR compliant blockchain usage can be expanded.

## V. CONTRIBUTIONS

In this work, we have outlined the compliance issues of blockchain with GDPR. We have discussed the areas that have been explored in prior literature. Furthermore, we have identified a number of research gaps. Therefore, our work has several implications for future research and practice. The implications are discussed below.

- i. The first and most important contribution of this work is to the best of our knowledge there is no previous systematic review on blockchain and GDPR compliance issues. Our paper synthesizes the findings in this research topic. This literature review work will entice the researchers to find out the unexplored areas of research for their contributions. Moreover, this work will help them to find blockchain and GDPR compliant and non-compliant articles at one place and their comprehensive summary. This will also be valuable for practitioners. GDPR has an impact on companies in European regions and also companies that deal with EU citizen data around the world. The summarized results (presented in Figure 3 and 4) can help practitioners find possible solutions to a particular GDPR challenge in their business case.
- ii. Second, this paper presents a comprehensive analysis of the GDPR articles that have been explored for compliance issues. We have seen that the right to be forgotten has been explored more and various directions have been proposed [e.g., 28-34]. Similarly, data processing and controlling role definitions have also been explored vastly [40-49]. In addition to that, privacy by design and their integration has been discussed [e.g., 30, 31, 35]. The concise representation of the articles explored facilitates the researchers to explore the less explored GDPR articles. Some of these less explored articles are related to data subject's consent management, data subjects' right, territorial scope, and defining joint controllers. Analyzing these issues will help to scale up the research scopes in the future.
- iii. Third, we have analyzed and provided a very detailed overview of the research areas that have been taken into consideration until now. For example, healthcare, identity management, and IoT are the most explored domains for compliance issues [e.g., 48, 62-66]. Some areas, for example, cryptocurrency, smart grid, smart and wearable devices, industrial data are unexplored. These areas involve and produce a large amount of personal data. Our observation from this analysis will help the potential researchers to gaze their sight onto those areas. As a result, possible compliance issues and state-of-the-art solutions can be found for those unexplored areas.
- iv. Finally, this article can be a guideline for learning about the GDPR compliance issues with IoT and blockchain-based industrial data. From our analysis, we have observed that IoT is a major area, where GDPR compliance issues are critical. Since IoT and blockchain both are essential building blocks of future industrial applications, findings from this work can help researchers, enthusiasts and other actors in building ethically sustainable digital services that follow GDPR recommendations.

## VI. CONCLUDING REMARKS

This SLR aimed to synthesize the prior works on the topic of GDPR compliant blockchains. We collected the related studies from two major databases and conducted an in-depth analysis of the studies. Our SLR clearly indicated that the research studies on this topic have been rising. We found that GDPR articles on data deletion and modification (Article 16 and 17) are among the most discussed compliance issues for blockchain. Furthermore, we have also observed that the role distribution of different actors as data controllers and processors have also been widely discussed in prior literature. We have identified eight research areas based on our thematic analysis. Among these themes, we observed IoT and healthcare domains are the most discussed research areas in prior literature. There are two major limitations of this SLR. First, we included only peer-reviewed articles in our study. This means that our approach excluded industry or practitioner reports, which might provide additional insights into the topic. Second, we have conducted our searches only in two databases. We have used citation chaining to find additional articles. Despite this, we might have missed some important articles to be considered in this SLR. This limitation can be addressed in future research by adding additional databases for searching articles.

**REFERENCES**

[1] Islam, AKM Najmul, Matti Mäntymäki, and Marja Turunen. "Why do blockchains split? An actor-network perspective on Bitcoin splits." Technological Forecasting and Social Change 148 (2019): 119743.

[2] Mohanta, Bhabendu Kumar, Soumyashree S. Panda, and Debasish Jena. "An overview of smart contract and use cases in blockchain technology." 2018 9th International Conference on Computing, Communication and Networking Technologies (ICCCNT). IEEE, 2018.

[3] Hjerppe, Kalle, Jukka Ruohonen, and Ville Leppänen. "The general data protection regulation: requirements, architectures, and constraints." 2019 IEEE 27th International Requirements Engineering Conference (RE). IEEE, 2019.

[4] "General Data Protection Regulation (GDPR) – Official Legal Text." https://gdpr-info.eu/ (accessed Dec. 28, 2020).

[5] "Blockchain and the General Data Protection Regulation",2019. Accessed: Dec. 27, 2020. [Online]. Available: https://www.europarl.europa.eu/RegData/etudes/STUD/2019/634445/EPRS_STU(2019)634445_EN.pdf

[6] Lambrinoudakis, Costas. "The general data protection regulation (GDPR) era: ten steps for compliance of data processors and data controllers." International Conference on Trust and Privacy in Digital Business. Springer, Cham, 2018.

[7] Sion, Laurens, et al. "An architectural view for data protection by design." 2019 IEEE International Conference on Software Architecture (ICSA). IEEE, 2019.

[8] Tandon, Anushree, et al. "Blockchain in healthcare: A systematic literature review, synthesizing framework and future research agenda." Computers in Industry 122 (2020): 103290.

[9] Koutli, Maria, et al. "Secure IoT e-Health applications using VICINITY framework and GDPR guidelines." 2019 15th International Conference on Distributed Computing in Sensor Systems (DCOSS). IEEE, 2019.

[10] Campanile, Lelio, et al. "Designing a GDPR compliant blockchain-based IoV distributed information tracking system." Information Processing & Management 58.3 (2021): 102511.

[11] Radinger-Peer, Wolfgang, and Bernhard Kolm. "A blockchain-driven approach to fulfill the gdpr recording requirements." Blockchain and Distributed Ledger Technology Use Cases. Springer, Cham, 2020. 133-148.

[12] B. Kitchenham, and S. Charters, "Guidelines for performing Systematic Literature Reviews in Software Engineering," 2007, Accessed: Dec. 27, 2020. [Online]. Available: http://citeseerx.ist.psu.edu/viewdoc/summary?doi=10.1.1.117.471.

[13] S. Nakamoto, "Bitcoin: A Peer-to-Peer Electronic Cash System | Satoshi Nakamoto Institute," Oct. 2008. Accessed: Dec. 28, 2020. [Online]. Available: https://nakamotoinstitute.org/bitcoin/.

[14] Sunyaev, Ali. "Distributed ledger technology." Internet Computing. Springer, Cham, 2020. 265-299.

[15] Mingxiao, Du, et al. "A review on consensus algorithm of blockchain." 2017 IEEE international conference on systems, man, and cybernetics (SMC). IEEE, 2017.

[16] Ismail, Leila, and Huned Materwala. "A review of blockchain architecture and consensus protocols: Use cases, challenges, and solutions." Symmetry 11.10 (2019): 1198.

[17] B. A. Tama, B. J. Kweka, Y. Park, and K. H. Rhee, "A critical review of blockchain and its current applications," in ICECOS 2017 - Proceeding of 2017 International Conference on Electrical Engineering and Computer Science: Sustaining the Cultural Heritage Toward the Smart Environment for Better Future, 2017, pp. 109–113, doi: 10.1109/ICECOS.2017.8167115.

[18] T. Ali Syed, A. Alzahrani, S. Jan, M. S. Siddiqui, A. Nadeem, and T. Alghamdi, "A Comparative Analysis of Blockchain Architecture and its Applications: Problems and Recommendations," IEEE Access, vol. 7, pp. 176838–176869, 2019, doi: 10.1109/ACCESS.2019.2957660.

[19] B. K. Mohanta, D. Jena, S. S. Panda, and S. Sobhanayak, "Blockchain technology: A survey on applications and security privacy Challenges," Internet of Things, vol. 8, p. 100107, Dec. 2019, doi: 10.1016/j.iot.2019.100107.

[20] Gupta, Suyash, Sajjad Rahnama, and Mohammad Sadoghi. "Permissioned blockchain through the looking glass: Architectural and implementation lessons learned." arXiv preprint arXiv:1911.09208 (2019).

[21] M. Lovell and M. A. Foy, "General Data Protection Regulation May 2018 (GDPR)," Bone Jt. 360, vol. 7, no. 4, pp. 41–42, Aug. 2018, doi: 10.1302/2048-0105.74.360622.

[22] J. Krystlik, "With GDPR, preparation is everything," Comput. Fraud Secur., vol. 2017, no. 6, pp. 5–8, Jun. 2017, doi: 10.1016/S1361-3723(17)30050-7.

[23] Mourby, Miranda, et al. "Are 'pseudonymised'data always personal data? Implications of the GDPR for administrative data research in the UK." Computer Law & Security Review 34.2 (2018): 222-233.

[24] G. Almeida Teixeira, M. Mira da Silva, and R. Pereira, "The critical success factors of GDPR implementation: a systematic literature review," Digital Policy, Regulation and Governance, vol. 21, no. 4. Emerald Group Publishing Ltd., pp. 402–418, Jun. 10, 2019, doi: 10.1108/DPRG-01-2019-0007.

[25] Hoofnagle, Chris Jay, Bart van der Sloot, and Frederik Zuiderveen Borgesius. "The European Union general data protection regulation: what it is and what it means." Information & Communications Technology Law 28.1 (2019): 65-98.

[26] Mantelero, Alessandro. "The EU Proposal for a General Data Protection Regulation and the roots of the 'right to be forgotten'."

Computer Law & Security Review 29.3 (2013): 229-235. doi: 10.1016/j.clsr.2013.03.010

[27] De Hert, Paul, and Vagelis Papakonstantinou. "The proposed data protection Regulation replacing Directive 95/46/EC: A sound system for the protection of individuals." Computer Law & Security Review 28.2 (2012): 130-142. doi: 10.1016/j.clsr.2012.01.011

[28] B. Varghese et al., "Realizing Edge Marketplaces: Challenges and Opportunities," IEEE Cloud Comput., vol. 5, no. 6, pp. 9–20, Nov. 2018, doi: 10.1109/MCC.2018.064181115. *

[29] F. Zemler and M. Westner, "Blockchain and GDPR: Application scenarios and compliance requirements," in PICMET 2019 - Portland International Conference on Management of Engineering and Technology: Technology Management in the World of Intelligent Systems, Proceedings, Aug. 2019, pp. 1–8, doi: 10.23919/PICMET.2019.8893923.

[30] A. Giannopoulou and V. Ferrari, "Distributed data protection and liability on blockchains," in Lecture Notes in Computer Science (including subseries Lecture Notes in Artificial Intelligence and Lecture Notes in Bioinformatics), Oct. 2019, vol. 11551 LNCS, pp. 203–211, doi: 10.1007/978-3-030-17705-8_17.

[31] W. L. Sim, H. N. Chua, and M. Tahir, "Blockchain for Identity Management: The Implications to Personal Data Protection," in 2019 IEEE Conference on Application, Information and Network Security (AINS), Nov. 2019, pp. 30–35, doi: 10.1109/AINS47559.2019.8968708.

[32] F. Casino, E. Politou, E. Alepis, and C. Patsakis, "Immutability and Decentralized Storage: An Analysis of Emerging Threats," IEEE Access, vol. 8, pp. 4737–4744, 2020, doi: 10.1109/ACCESS.2019.2962017.

[33] Ł. Wyciślik and E. Marcinkowska, "Tracking of clinical documentation based on the blockchain technology—A Polish case study," Sustain., vol. 12, no. 22, pp. 1–14, Nov. 2020, doi: 10.3390/su12229517.

[34] N. B. Truong, K. Sun, G. M. Lee, and Y. Guo, "GDPR-Compliant Personal Data Management: A Blockchain-based Solution," IEEE Trans. Inf. Forensics Secur., vol. 15, pp. 1746–1761, Apr. 2019, doi: 10.1109/TIFS.2019.2903659.

[35] S. Ma et al., "Nudging data privacy management of open banking based on blockchain," in Proceedings - 2018 15th International Symposium on Pervasive Systems, Algorithms and Networks, I-SPAN 2018, Oct. 2019, pp. 72–79, doi: 10.1109/I-SPAN.2018.00021.

[36] E. Kadena and P. Holicza, "Security Issues in the Blockchain(ed) World," in 18th IEEE International Symposium on Computational Intelligence and Informatics, CINTI 2018 - Proceedings, Nov. 2018, pp. 211–215, doi: 10.1109/CINTI.2018.8928212. *

[37] C. Damian, I. Lazar, D.-G. Visoiu, S. Romanescu, and L. Alboaie, "Applying Blockchain Technologies in Funding of Electrical Engineering Industry Applications," in 2019 International Conference on Electromechanical and Energy Systems (SIELMEN), Oct. 2019, pp. 1–5, doi: 10.1109/SIELMEN.2019.8905896.

[38] N. Al-Zaben, M. M. H. Onik, J. Yang, N. Y. Lee, and C. S. Kim, "General Data Protection Regulation Complied Blockchain Architecture for Personally Identifiable Information Management," in Proceedings - 2018 International Conference on Computing, Electronics and Communications Engineering, iCCECE 2018, Aug. 2019, pp. 77–82, doi: 10.1109/iCCECOME.2018.8658586.

[39] W.-C. Huang, L.-Y. Yeh, and J.-L. Huang, "A Monitorable Peer-to-Peer File Sharing Mechanism," in 2019 20th Asia-Pacific Network Operations and Management Symposium (APNOMS), Sep. 2019, pp. 1–4, doi: 10.23919/APNOMS.2019.8892963. *

[40] R. Herian, "Blockchain, GDPR, and fantasies of data sovereignty," Law, Innov. Technol., vol. 12, no. 1, pp. 156–174, Jan. 2020, doi: 10.1080/17579961.2020.1727094.

[41] T. Buocz, T. Ehrke-Rabel, E. Hödl, and I. Eisenberger, "Bitcoin and the GDPR: Allocating responsibility in distributed networks," Comput. Law Secur. Rev., vol. 35, no. 2, pp. 182–198, Apr. 2019, doi: 10.1016/j.clsr.2018.12.003.

[42] A. Bayle, M. Koscina, D. Manset, and O. Perez-Kempner, "When Blockchain Meets the Right to Be Forgotten: Technology versus Law in the Healthcare Industry," in Proceedings - 2018 IEEE/WIC/ACM International Conference on Web Intelligence, WI 2018, Dec. 2019, pp. 788–792, doi: 10.1109/WI.2018.00133. *

[43] A. Giannopoulou, "Data protection compliance challenges for self-sovereign identity," in Advances in Intelligent Systems and Computing, Jun. 2020, vol. 1238 AISC, pp. 91–100, doi: 10.1007/978-3-030-52535-4_10.

[44] A. Jambert, "Blockchain and the GDPR: A data protection authority point of view," in Lecture Notes in Computer Science (including subseries Lecture Notes in Artificial Intelligence and Lecture Notes in Bioinformatics), Dec. 2019, vol. 11469 LNCS, pp. 3–6, doi: 10.1007/978-3-030-20074-9_1.

[45] R. Dutta, A. Das, A. Dey, and S. Bhattacharya, "Blockchain vs GDPR in Collaborative Data Governance," in Lecture Notes in Computer Science (including subseries Lecture Notes in Artificial Intelligence and Lecture Notes in Bioinformatics), vol. 12341 LNCS, Springer Science and Business Media Deutschland GmbH, 2020, pp. 81–92.

[46] G. Kondova and J. Erbguth, "Self-sovereign identity on public blockchains and the GDPR," in Proceedings of the ACM Symposium on Applied Computing, Mar. 2020, pp. 342–345, doi: 10.1145/3341105.3374066.

[47] U. Tatar, Y. Gokce, and B. Nussbaum, "Law versus technology: Blockchain, GDPR, and tough tradeoffs," Comput. Law Secur. Rev., vol. 38, p. 105454, Sep. 2020, doi: 10.1016/j.clsr.2020.105454.

[48] D. Hofman, V. L. Lemieux, A. Joo, and D. A. Batista, "'The margin between the edge of the world and infinite possibility,'" Rec. Manag. J., vol. 29, no. 1/2, pp. 240–257, Mar. 2019, doi: 10.1108/RMJ-12-2018-0045.

[49] L. F. M. Ramos and J. M. C. Silva, "Privacy and data protection concerns regarding the use of blockchains in smart cities," in ACM International Conference Proceeding Series, 2019, vol. Part F1481, pp. 342–347, doi: 10.1145/3326365.3326410.

[50] D. Schmelz, K. Pinter, J. Brottrager, P. Niemeier, R. Lamber, and T. Grechenig, "Securing the rights of data subjects with blockchain technology," in Proceedings - 3rd International Conference on Information and Computer Technologies, ICICT 2020, Mar. 2020, pp. 284–288, doi: 10.1109/ICICT50521.2020.00050.

[51] N.-Y. Lee, J. Yang, M. M. H. Onik, and C.-S. Kim, "Modifiable Public Blockchains Using Truncated Hashing and Sidechains," IEEE Access, vol. 7, pp. 173571–173582, 2019, doi: 10.1109/ACCESS.2019.2956628.

[52] J. Ahmed, S. Yildirim, M. Nowostaki, R. Ramachandra, O. Elezaj, and M. Abomohara, "GDPR compliant consent driven data protection in online social networks: A blockchain-based approach," in Proceedings - 3rd International Conference on Information and Computer Technologies, ICICT 2020, Mar. 2020, pp. 307–312, doi: 10.1109/ICICT50521.2020.00054.

[53] R. Belen Saglam, C. B. Aslan, S. Li, L. Dickson, and G. Pogrebna, "A Data-Driven Analysis of Blockchain Systems' Public Online Communications on GDPR," in 2020 IEEE International Conference on Decentralized Applications and Infrastructures (DAPPS), Aug. 2020, pp. 22–31, doi: 10.1109/DAPPS49028.2020.00003.

[54] G. P. Freund, P. B. Fagundes, and D. D. J. de Macedo, "An Analysis of Blockchain and GDPR under the Data Lifecycle Perspective," Mob. Networks Appl., pp. 1–11, Aug. 2020, doi: 10.1007/s11036-020-01646-9.

[55] G. S. Aujla et al., "COM-PACE: Compliance-Aware Cloud Application Engineering Using Blockchain," IEEE Internet Comput., vol. 24, no. 5, pp. 45–53, Sep. 2020, doi: 10.1109/MIC.2020.3014484.

[56] M. Pedrosa, A. Zúquete, and C. Costa, "RAIAP: renewable authentication on isolated anonymous profiles: A GDPR compliant self-sovereign architecture for distributed systems," Peer-to-Peer Netw. Appl., vol. 13, no. 5, pp. 1577–1599, Sep. 2020, doi: 10.1007/s12083-020-00914-5.

[57] M. Koscina, D. Manset, C. Negri, and O. P. Kempner, "Enabling trust in healthcare data exchange with a federated blockchain-based architecture," in Proceedings - 2019 IEEE/WIC/ACM International Conference on Web Intelligence Workshops, WI 2019 Companion, Oct. 2019, pp. 231–237, doi: 10.1145/3358695.3360897.

[58] O. P. Stan and L. Miclea, "New era for technology in healthcare powered by GDPR and blockchain," in IFMBE Proceedings, 2019, vol. 71, pp. 311–317, doi: 10.1007/978-981-13-6207-1_49.

[59] M. N. Asghar, N. Kanwal, B. Lee, M. Fleury, M. Herbst, and Y. Qiao, "Visual Surveillance Within the EU General Data Protection Regulation: A Technology Perspective," IEEE Access, vol. 7, pp. 111709–111726, Aug. 2019, doi: 10.1109/access.2019.2934226. *

[60] E. Politou, F. Casino, E. Alepis, and C. Patsakis, "Blockchain Mutability: Challenges and Proposed Solutions," IEEE Trans. Emerg. Top. Comput., pp. 1–1, 2020, doi: 10.1109/TETC.2019.2949510.

[61] I. Makhdoom, I. Zhou, M. Abolhasan, J. Lipman, and W. Ni, "PrivySharing: A blockchain-based framework for privacy-preserving and secure data sharing in smart cities," Comput. Secur., vol. 88, p. 101653, Jan. 2020, doi: 10.1016/j.cose.2019.101653

[62] M. Barati, I. Petri, and O. F. Rana, "Developing GDPR compliant user data policies for internet of things," in UCC 2019 - Proceedings of the 12th IEEE/ACM International Conference on Utility and Cloud Computing, Dec. 2019, pp. 133–141, doi: 10.1145/3344341.3368812.

[63] D. Billard and B. Bartolomei, "Digital Forensics and Privacy-by-Design: Example in a Blockchain-Based Dynamic Navigation System," in Lecture Notes in Computer Science (including subseries Lecture Notes in Artificial Intelligence and Lecture Notes in Bioinformatics), vol. 11498 LNCS, Springer Verlag, 2019, pp. 151–160.

[64] N. Fabiano, "The Internet of Things ecosystem: The blockchain and privacy issues. The challenge for a global privacy standard," in 2017 International Conference on Internet of Things for the Global Community (IoTGC), Jul. 2017, pp. 1–7, doi: 10.1109/IoTGC.2017.8008970. *

[65] X. Zheng, R. R. Mukkamala, R. Vatrapu, and J. Ordieres-Mere, "Blockchain-based Personal Health Data Sharing System Using Cloud Storage," in 2018 IEEE 20th International Conference on e-Health Networking, Applications and Services (Healthcom), Sep. 2018, pp. 1–6, doi: 10.1109/HealthCom.2018.8531125.

[66] F. Mohammadi, A. Panou, C. Ntantogian, E. Karapistoli, E. Panaousis, and C. Xenakis, "CUREX: Secure and private health data exchange," in Proceedings - 2019 IEEE/WIC/ACM International Conference on Web Intelligence Workshops, WI 2019 Companion, Oct. 2019, pp. 263–268, doi: 10.1145/3358695.3361753.

[67] A. Rahman, M. K. Nasir, Z. Rahman, A. Mosavi, S. Shahab, and B. Minaei-Bidgoli, "DistBlockBuilding: A Distributed Blockchain-Based SDN-IoT Network for Smart Building Management," IEEE Access, vol. 8, pp. 140008–140018, 2020, doi: 10.1109/ACCESS.2020.3012435.

[68] A. Dorri, S. S. Kanhere, R. Jurdak, and P. Gauravaram, "Blockchain for IoT security and privacy: The case study of a smart home," in 2017 IEEE International Conference on Pervasive Computing and Communications Workshops, PerCom Workshops 2017, May 2017, pp. 618–623, doi: 10.1109/PERCOMW.2017.7917634.

[69] H. Li and D. Han, "EduRSS: A Blockchain-Based Educational Records Secure Storage and Sharing Scheme," IEEE Access, vol. 7, pp. 179273–179289, 2019, doi: 10.1109/ACCESS.2019.2956157.

[70] H. Yumna, M. M. Khan, M. Ikram, and S. Ilyas, "Use of Blockchain in Education: A Systematic Literature Review," in Lecture Notes in Computer Science (including subseries Lecture Notes in Artificial Intelligence and Lecture Notes in Bioinformatics), Apr. 2019, vol. 11432 LNAI, pp. 191–202, doi: 10.1007/978-3-030-14802-7_17.

[71] Islam, Iyolita, et al. "A critical review of concepts, benefits, and pitfalls of blockchain technology using concept map." IEEE Access 8 (2020): 68333-68341.

[72] Yli-Huumo, J., Ko, D., Choi, S., Park, S., & Smolander, K. (2016). Where is current research on blockchain technology?—a systematic review. PloS one, 11(10), e0163477.

[73] Seebacher, Stefan, and Ronny Schüritz. "Blockchain technology as an enabler of service systems: A structured literature review." International Conference on Exploring Services Science. Springer, Cham, 2017.

[74] Tama, Bayu Adhi, et al. "A critical review of blockchain and its current applications." 2017 International Conference on Electrical Engineering and Computer Science (ICECOS). IEEE, 2017.

[75] Casino, Fran, Thomas K. Dasaklis, and Constantinos Patsakis. "A systematic literature review of blockchain-based applications: current status, classification and open issues." Telematics and informatics 36 (2019): 55-81.

[76] Pournader, Mehrdokht, et al. "Blockchain applications in supply chains, transport and logistics: a systematic review of the literature." International Journal of Production Research 58.7 (2020): 2063-2081.

[77] Tang, Yong, et al. "Ethics of blockchain." Information Technology & People (2019).

[78] Dierksmeier, Claus, and Peter Seele. "Blockchain and business ethics." Business Ethics: A European Review 29.2 (2020): 348-359.

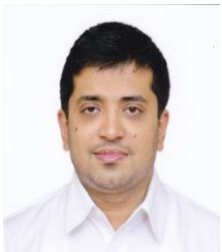

AKM Bahalul Haque is a Lecturer at the Dept. of Electrical and Computer Engineering at North South University. He teaches undergraduate courses and supervises final theses/projects. Recently he has been accepted for the position of Junior Researcher at LUT University Finland. His and his teams works have been published in IEEE conference proceedings, journals, and book chapters. His research area includes blockchain, IoT, data privacy and protection, and human-computer interaction.

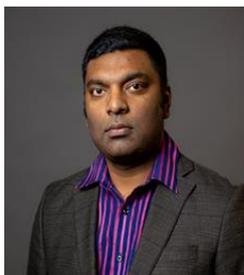

A.K.M. Najmul Islam is an Adjunct Professor at Tampere University, Finland. He is an Associate Professor at LUT University, Finland. Dr. Islam holds a PhD (Information Systems) from the University of Turku, Finland and an M.Sc. (Eng.) from Tampere University of Technology, Finland. His research focuses on Human Cantered Computing. His research has been published in top outlets such as IEEE Access, European Journal of Information Systems, Information Systems Journal, Journal of Strategic Information Systems, Technological Forecasting and Social Change, Computers in Human Behavior, Internet Research, Computers & Education, Journal of Medical Internet Research, Information Technology & People, Telematics & Informatics, Journal of Retailing and Consumer Research, Communications of the AIS, Journal of Information Systems Education, AIS Transaction on Human-Computer Interaction, and Behaviour & Information Technology, among others.

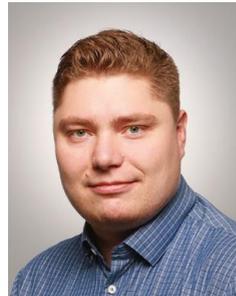

Sami Hyrynsalmi is a nerd who has always enjoyed playing with computers and programs. He currently works as an Associate Professor of Software Engineering in LUT University's Lahti Campus. He graduated as a Master of Science in Technology (Software Engineering) in 2009 and Doctor of Science in Technology (Information and Communication Technologies) in 2014. During his career, he has worked for University of Turku, Turku School of Economics, Tampere University of Technology as well as for Tampere University. After defending his thesis, he has focused on various themes from security to software measurement, and from gamification ethics to business ecosystems and software product management. He has authored over 20 journal articles and more than 80 conference and workshop papers. His and his colleagues work has been published in journal such as Information and Software Technology, Journal of Systems and Software, Computers in Human Behavior, Telematics and Informatics and Electronic Markets.

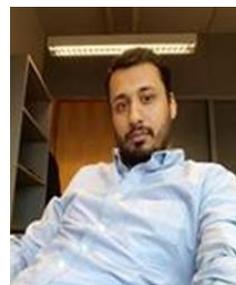

Dr. BILAL NAQVI is a post-doctoral researcher at LUT University, Finland. His main research focus is on studying the interdependencies, conflicts, and trade-offs between usability and security. Besides this, he is also interested in other issues related to cyber security and digitalization. Previously, he has received B.E Computer Software Engineering and M.S. Information Security degrees from the National University of Sciences and Technology, NUST, Pakistan. Dr. Naqvi is also involved in teaching different courses of Software Engineering programs at LUT University.

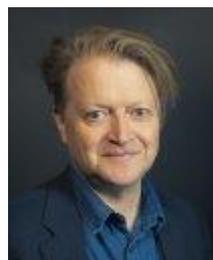

Kari Smolander is professor and Head of Software Engineering Department at LUT University, Finland, and an adjunct professor in Aalto University, Finland. In addition to his long teaching and research experience, he has also an earlier industry career. His works have been published in journals such as Information and Software Technology, Journal of Systems and Software and European Journal of Information Systems. His current research interests include change in software and systems development practices and software development organizations.